\documentclass[amsfonts,preprint,aps]{revtex4}

\begin{document}

\title{Holography from loop quantum gravity}

\author{ 
Rodolfo Gambini$^{1}$,
Jorge Pullin$^{2}$}
\affiliation {
1. Instituto de F\'{\i}sica, Facultad de Ciencias, 
Igu\'a 4225, esq. Mataojo, Montevideo, Uruguay rgambini@fisica.edu.uy \\
2. Department of Physics and Astronomy, Louisiana State University,
Baton Rouge, LA 70803-4001 pullin@lsu.edu}

\begin{abstract}
We show that holography follows directly from the basic
structure of spherically symmetric loop quantum gravity. The result
is not dependent on detailed assumptions about the dynamics of 
the theory being considered. It ties strongly the amount
of information contained in a region of space to the tight 
mathematical underpinnings of loop quantum geometry.
\end{abstract}

\maketitle

Physical principles usually represent facts partially collected by
observation that end up guiding the development of physical
theories. When the underlying theories are completely understood in
detail, the principles can be explained as consequences of
the theory they guided to create. The holographic principle has guided
the construction of some of the leading physical theories of space
time in the last few years. In a nutshell, holography establishes a
limit to the amount of information contained in a space-time region
\cite{bousso}. In its simplest form, for spherical symmetry and weak
gravity the principle establishes that the entropy of a region of
space is limited by the area surrounding it and was first formulated
by t'Hooft and Susskind \cite{thooftsusskind}. Any
successful theory of quantum gravity that incorporates holography
should be able to derive it as a consequence of its framework. We
would like to argue that holography does indeed follow from the
framework of loop quantum gravity in spherical symmetry and that the
result is robust: it does not depend on the details of the dynamics
of the theory nor the type of matter included but rather on its
kinematical structure and elementary dynamical considerations independent
of the details of the Hamiltonian. That holography in its simple and
straightforward {\em spatial} form is materialized in the spherical
case is appropriate, since it is known that in non-spherical cases
more care is needed (in particular involving spatiotemporal regions)
in its definition in order not to run into counterexamples 
(see \cite{bousso} for a discussion of this point).

The argument we will present can be summarized as follows:
 holography follows from the dependence of the volume
operator in spherical loop quantum gravity on the radial distance, yielding an
uncertainty in the determination of volumes that grows radially. Such
a dependence for the uncertainty of spatial measurements had already
been postulated in heuristic treatments relating limitations of
space-time measurements to holography by Ng
\cite{ng} and with alternative reasonings by Ng and Lloyd
\cite{lloyd}). In this article we show that such a dependence can be
derived from the kinematical structure of spherical loop quantum
gravity.

We consider spherically symmetric loop quantum gravity. Its kinematic
structure
is well established and was discussed in detail by Bojowald and
Swiderski \cite{boswi}. There is only one non-trivial spatial
direction (the radial) which we call $x$ since it is not necessarily
parameterized by the usual radial coordinate. 
The only non-trivial components of the metric
are given in terms of the Ashtekar triads by 
$g_{xx}= {(E^\varphi)^2}/{|E^x|}$ and $g_{\theta\theta} = |E^x|$.
It is convenient to gauge fix the radial direction (for further details
see \cite{cagapu}) to the usual Schwarzschild coordinate
$E^x-(x+2M)^2=0$ with $M$ the mass of the space-time and
$x\in {\mathbb R}^+$ and the horizon at $x=0$.
In terms of these variables the volume of shell of radial interval $I$ 
is given by,
\begin{equation}
  V(I) = 4\pi \int dx |E^\varphi(x)| (x+2M).
\end{equation}

One can introduce a loop representation. The ``loops'' consist of
intervals in the radial direction and the ``vertices'' are the ends of
the intervals. Since one has gauge fixed the radial direction,
variables behave as scalars and the loop representation resembles the
one introduced for loop quantum cosmology, except that there is one
such variable per vertex. Essentially the variables conjugate to the
triads are represented through their exponentiated form and depend on
a parameter that appears in the exponent per vertex, $\mu_v$.  States
in the loop representation are labeled by collections of real valued
parameters $|\vec{\mu}\rangle=|\mu_1,\ldots,\mu_n\rangle$. The volume
operator can be readily quantized to give,
\begin{equation}
  \hat{V}(I) |\vec{\mu}\rangle = \sum_{v \in I} 4\pi |\mu_v| (x_v+2M) 
\gamma \ell_{\rm Planck}^2 |\vec{\mu}\rangle
\end{equation}
where $\gamma$ is the Immirzi parameter. On this kinematical arena
one will have to build the dynamics of the theory of interest, be it 
general relativity or some other theory, including possible matter
couplings. In order to build the dynamics we need the action of the
exponentiated variable that plays the role in this reduced context of
the holonomies of the loop representation. As is customary, the action
of these elementary operators in the loop representation is particularly
simple,
\begin{equation}
\hat{h}_\varphi(v_i,\rho) | g, \vec{\mu}> = | g,\mu_{v_1},
\ldots,\mu_{v_i}+\rho,\ldots>.
\end{equation}

Any candidate for a Hamiltonian of the theory one may wish to build
will involve the action of the elementary operator $\hat{h}$. Usually
the Hamiltonian one starts with will be a function of the connection,
which can be expressed as a limit of the elementary operator when
$\rho\to 0$. In the loop representation the unique measure that 
arises \cite{lost} prevents one from taking the limit and one has
to take a minimum value for $\rho$. 

The crucial observation is to note that the action of the elementary
operator, on which all possible Hamiltonians will be based, takes an
eigenstate of the volume operator and produces a new eigenstate where
the volume has increased by 
\begin{equation}\label{eqng}
\Delta V = 4 \pi 2\gamma \rho
\ell_{\rm Planck}^2 (x+2 M)
\end{equation}
where $(x+2M)$ is the usual Schwarzschild
coordinate.  If we take for $\rho$ the minimum value possible, this
quantity plays a role of minimum volume for the models we are
considering.  To get a handle on the minimum possible value of $\rho$,
we use a reasoning similar to the one used in loop quantum
cosmology. This estimate comes from the fact that in the full theory
areas have a minimum quantum and this leaves an imprint on the
variables of the symmetry reduced models. The estimate 
\cite{estimate} 
from loop
quantum cosmology \cite{asbole} is $\rho=\sqrt{3}/4$.  We can then
evaluate the number $\Delta N$ of such elementary volumes in a shell
of width $\Delta x$ (in the asymptotic region where we assume the
metric is flat, otherwise one would have to add a finite correction and
substitute $x$ by $x+2M$),
\begin{equation}
\Delta  N = \frac{4 \pi x^2 \Delta x}{4\pi 2 \gamma \rho x 
\ell_{\rm Planck}^2}=
\frac{x \Delta x}{2\gamma \rho \ell^2_{\rm Planck}},
\end{equation}
We can now compute the entropy in a shell as the one discussed,
\begin{equation}
\Delta S = \nu_v \frac{x \Delta x}{2\gamma \rho \ell^2_{\rm Planck}},
\end{equation}
where $\nu_v$ is the mean entropy per unit volume.

The Immirzi parameter $\gamma = c_A /(\pi \sqrt{3})$ with $c_A$ is a
quantity that is to be determined by comparing physical predictions of
the theory with reality. For instance, calculations of the entropy of
black holes suggests it is of order unity.  We can therefore write for
the entropy of an infinitesimal shell,
\begin{equation}
\Delta S =\frac{\nu_v}{c_A} 4\pi \frac{x \Delta x}{2 \ell_{\rm Planck}^2},
\end{equation}
so for a finite shell of inner radius $r_a$ and outer radius $r_b$ one
would have,
\begin{equation}
S =\frac{\nu_v}{c_A} 4\pi \frac{r_b^2-r_a^2}{4 \ell_{\rm Planck}^2}=
\frac{1}{4}\frac{\nu_v}{c_A} \frac{\Sigma_b-\Sigma_a}{\ell^2_{\rm Planck}},
\end{equation}
where $\Sigma$ is the area of the shell of the given radius
which implies that the entropy is proportional to the area. This quantity is a
upper bound for the entropy, since we have used the minimum value
of $\rho$, obviously choosing larger $\rho$'s one would obtain lower values
for $S$.

We have therefore established that the kinematical structure of loop
quantum gravity in spherical symmetry implies the holographic principle
irrespective of the dynamics of the theory being studied. It is
therefore a very general result. It stems from the fact that the
elementary volume that any dynamical operator may involve goes as
$ x \ell_{\rm Planck}^2$ (as suggested by previous heuristic estimates
\cite{ng}). We have assumed a finite amount of information per
elementary volume, as is usually argued in this context \cite{bousso}.
This implies that the information in a spatial region is bounded by
the area, contrary to what happens if one assumes the elementary
volume goes as $\ell_{\rm Planck}^3$. This is usually justified by
thinking that the fields are collections of harmonic oscillators and
the energy in each oscillator is bounded by the Planck energy
and therefore has a finite number of states. Although a complete
quantum gravity analysis has not been done, studies of the harmonic
oscillator \cite{corichi} and of linearized gravity \cite{madhavan}
suggest that this bound is even tighter in loop quantum gravity.

Holography is therefore naturally built into the elementary framework
of loop quantum gravity with spherical symmetry. The calculation we
showed also implies for the first time a derivation from first
principles of equation (\ref{eqng}) which had been heuristically
proposed \cite{ng} as a fundamental limit on the measurement of space
and time and the ultimate limits of computability in nature and which
may even be tested observationally in the near future in astronomical
settings \cite{ng2}. We can therefore consider that we have taken the
first steps to unravel the mystery of holography using some of the
most well established elements of traditional canonical quantization.
It is yet to be seen if these results are only a coincidence in the 
spherical case or if they can be found as well in more general settings.

We wish to thank Abhay Ashtekar for detailed comments about an earlier
version of this essay and Martin Bojowald, Parampreet Singh and Thomas
Thiemann for discussions.  We also wish to thank the Kavli Institute
for Theoretical Physics at the University of California at Santa
Barbara for hospitality.  This work was supported in part by grant
NSF-PHY-0554793, PHY-0551164, funds of the Hearne Institute for
Theoretical Physics and CCT-LSU and Pedeciba (Uruguay).

\end{document}